\documentclass[journal,twoside,web]{ieeecolor}
\usepackage{generic}
\usepackage{cite}
\usepackage{amsmath,amssymb,amsfonts}
\usepackage{algorithmic}
\usepackage{graphicx}
\usepackage{algorithm,algorithmic}
\usepackage{hyperref}
\hypersetup{hidelinks=true}
\usepackage{textcomp}
\usepackage{placeins}
\def\BibTeX{{\rm B\kern-.05em{\sc i\kern-.025em b}\kern-.08em
    T\kern-.1667em\lower.7ex\hbox{E}\kern-.125emX}}
\begin{document}
\title{SPG-CDENet: Spatial Prior-Guided Cross Dual Encoder Network for Multi-Organ Segmentation}
\author{Xizhi Tian, Changjun Zhou$^*$, and Yulin. Yang$^*$
\thanks{Xizhi Tian is with the School of Computer Science and Technology, Zhejiang Normal University, Jinhua, China. e-mail: tianxizhi0@gmail.com.}
\thanks{Changjun Zhou is with the School of Computer Science and Technology, Zhejiang Normal University, Jinhua, China. e-mail: zhouchangjun@zjnu.edu.cn.}
\thanks{Yulin Yang is with Metaverse and Artificial Intelligence Institute of Wenzhou University, Wenzhou, China. e-mail: yangyulin1991@gmail.com.}
\thanks{*Corresponding author.}}

\maketitle

\begin{abstract}
Multi-organ segmentation is a critical task in computer-aided diagnosis. While recent deep learning methods have achieved remarkable success in image segmentation, huge variations in organ size and shape challenge their effectiveness in multi-organ segmentation. To address these challenges, we propose a Spatial Prior-Guided Cross Dual Encoder Network (SPG-CDENet), a novel two-stage segmentation paradigm designed to improve multi-organ segmentation accuracy. Our SPG-CDENet consists of two key components: an spatial prior network and a cross dual encoder network. The prior network generates coarse localization maps that delineate the approximate ROI, serving as spatial guidance for the dual encoder network. The cross dual encoder network comprises four essential components: a global encoder, an local encoder, a symmetric cross-attention module, and a flow-based decoder. The global encoder captures global semantic features from the entire image, while the local encoder focuses on features from the prior network. To enhance the interaction between the global and local encoders, a symmetric cross-attention module is proposed across all layers of the encoders to fuse and refine features. Furthermore, the flow-based decoder directly propagates high-level semantic features from the final encoder layer to all decoder layers, maximizing feature preservation and utilization. Extensive qualitative and quantitative experiments on two public datasets—the Automated Cardiac Diagnosis Challenge and Synapse Multi-Organ CT—demonstrate the superior performance of SPG-CDENet compared to existing segmentation methods. Specifically, SPG-CDENet achieves a 95\% Hausdorff Distance of 12.75(Synapse) and a Dice Similarity Coefficient of 94.25\%, 85.97\%. Ablation studies further validate the effectiveness of the proposed modules in improving segmentation accuracy.
\end{abstract}

\begin{IEEEkeywords}
% Enter key words or phrases in alphabetical order, separated by commas. Using the IEEE Thesaurus can help you find the best standardized keywords to fit your article. Use the thesaurus access request form for free access to the IEEE Thesaurus: \underline{https://www.ieee.org/publications/services/thesaurus-acce}\\ % \underline{ss-page.com.}
Multi-organ Segmentation; 
Spatial Prior Network; 
Cross Dual Encoder Network; 
Symmetric Cross-Attention Module;  

 % \underline{ss-page.com.}
\end{IEEEkeywords}

\section{Introduction}
\label{sec:introduction}
\IEEEPARstart{M}{ulti-organ} segmentation is a foundational task in medical image analysis, focused on precisely delineating organ boundaries in medical images. It plays a critical role in computer-aided diagnosis, supporting applications such as disease detection, treatment planning, and surgical navigation. With the progress of deep learning, numerous models \cite{ronneberger2015unet, huang2022missformer, cao2021swinunet, sun2024paratranscnn, song2022global, ai2025sama, yang2024segmentation} have emerged to improve the accuracy and efficiency of multi-organ segmentation.

\begin{figure*}[t!]
    \centering
    \includegraphics[width=1\textwidth]{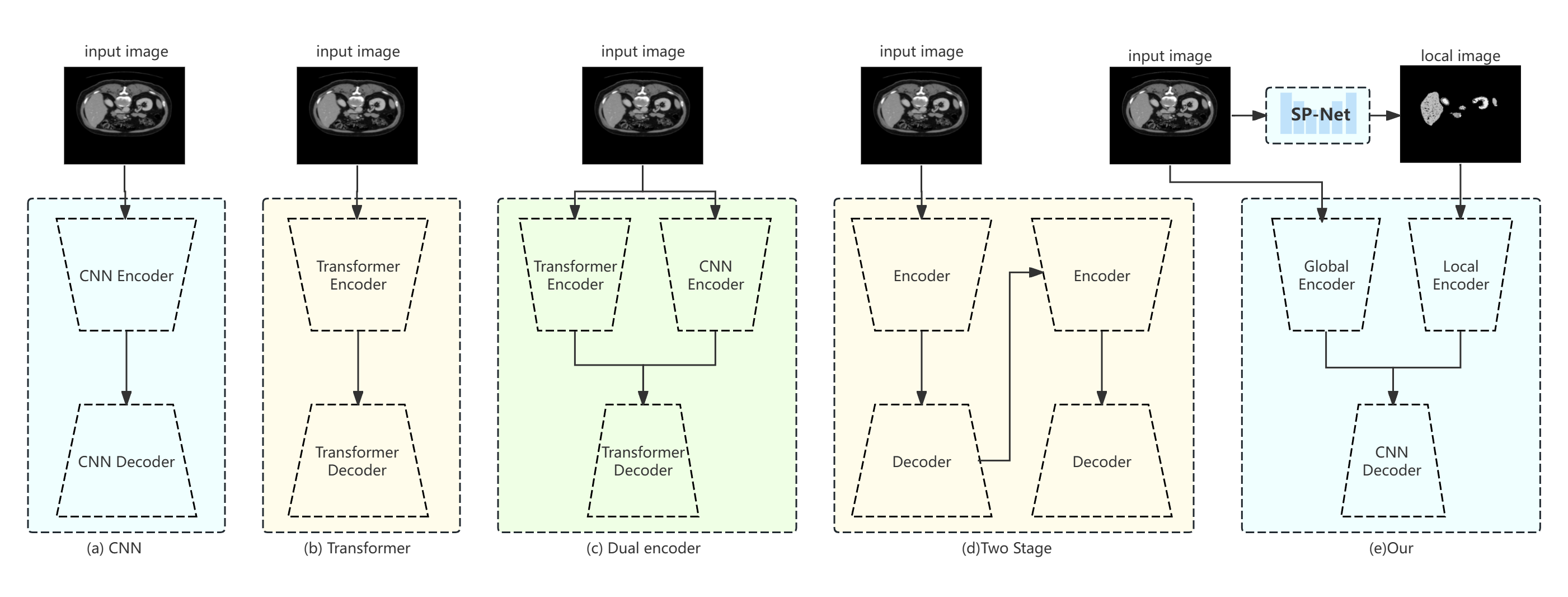}
    \caption{Conceptual comparison of the three most popular models used for medical image segmentation, where (a) classical U-Net\cite{ronneberger2015unet}; (b) Swin U-Net\cite{cao2021swinunet}; (c) ParaTransCNN encoder\cite{sun2024paratranscnn}; (d)Hierarchical FCN\cite{roth2017hierarchical}; (e)Our SPG-CDENet}
    \label{fig:enter-label}
\end{figure*}
U-Net-based image segmentation models have garnered significant attention due to their exceptional performance across various tasks. These models can be broadly classified into two primary categories based on the approach used to extract semantic features: CNN-based methods (Fig. 1(a)) and Transformer-based methods (Fig. 1(b)). Classic CNN-based architectures \cite{zhou2018unetplusplus, alom2019recurrent, li2023sunet} have demonstrated remarkable success in diverse image segmentation tasks. However, CNNs are inherently limited by their local receptive fields, which restrict their ability to capture long-range dependencies, often leading to suboptimal segmentation accuracy, particularly in complex or contextually rich images. To overcome these limitations, Transformer-based methods \cite{cao2021swinunet, chen2021transunet, huang2022missformer, wang2022Mixed} have been introduced for image segmentation. These models utilize self-attention mechanisms to effectively model global contextual information, thereby improving segmentation performance. For instance, MissFormer \cite{huang2022missformer} incorporates a multi-scale information integration module to better capture features across various spatial scales and levels of detail. Despite their strengths in global context modeling, Transformer-based models often struggle with capturing fine-grained local details, which can lead to blurred object boundaries, particularly in segmentation tasks requiring high spatial precision \cite{yu2024multiview}.

Although U-Net and its variants have achieved convincing performance across a wide range of image segmentation tasks, their effectiveness in multi-organ segmentation remains limited due to the inherent complexity of abdominal medical imaging. First, unlike natural images, abdominal scans are typically grayscale with low contrast, leading to ambiguous boundaries between organs and the surrounding background, as well as between adjacent organs with similar intensity and texture. These visual ambiguities significantly hinder the model's ability to achieve precise boundary delineation. Second, abdominal organs exhibit substantial inter-patient variability in appearance, including differences in shape, size, and texture. The substantial inter-patient variability in organ appearance significantly impairs model generalization, may resulting in prediction inconsistency, and reduced robustness on anatomically atypical cases. To address the first challenge, some studies \cite{zhang2025sacnet,  sun2024paratranscnn} have designed more complex network architectures to increase the receptive field, enabling the capture of finer-grained semantic features and thereby improving boundary segmentation accuracy. ParaTransCNN\cite{sun2024paratranscnn} incorporates an additional encoder path to improve the segmentation precision of complex anatomical boundaries, as shown in Fig.1 (C). Furthermore, to tackle the second challenge, several approaches \cite{huang2021atlas,zhong2025pgsam,wen2024dclnet} have introduced prior knowledge networks to localize organ regions in multi-organ segmentation tasks, achieving more precise segmentation results. For instance, PG-SAM\cite{zhong2025pgsam} introduces fine-grained semantic priors generated by a medical large language model to guide the Segment Anything Model toward more accurate multi-organ segmentation in medical images. Meanwhile, DCL‑Net\cite{wen2024dclnet} proposes a Dual‑Contrastive Learning framework wherein Stage-1 produces a mask prior to guide Stage-2’s organ-aware local refinement, as shown in Fig.1 (D).

Different from previous studies, we propose an Prior-Guided Cross Dual Encoder Network (SPG-CDENet) for more accurate multi-organ segmentation. As illustrated in Fig.1(e), SPG-CDENet includes two parts: a spatial prior network (SP-Net) and a crossing dual encoder network (CDE-Net). SP-Net employs a pre-trained multi-organ segmentation model followed by a post-processing module to localize region of interest(ROI) in the input image. These localized regions serve as spatial priors that guide the subsequent segmentation stage. To make fuller use of above localized regions, we device a crossing dual encoder network. CDE-Net consists of four parts: a global encoder, a local encoder,a symmetric cross-attention module and a flow-based decoder. The global encoder is responsible for extracting holistic features, while the local encoder focuses on the local features produced by the prior network. The symmetric cross-attention module integrates global and local features. The flow-based decoder propagates high-level semantic information to each layer of the decoder. Our contributions are summarized as follows:
\begin{itemize}
\item We propose a Prior-Guided Cross Dual Encoder Network, which consists of a spatial prior network and a crossing dual-encoder architecture. It can address the low boundary accuracy caused by background-interest similarity and the limited generalization due to large inter-interest variability. The effectiveness of our proposed model is validated through both qualitative and quantitative experiments.
\item We design a spatial prior network that utilizes a pre-trained model to obtain the prior location information. Experimental results demonstrate that our model exhibits strong robustness and generalization across different pre-trained models.
% Experimental results show that selecting pre-trained models with different levels of performance can accelerate the convergence of the overall framework, improve segmentation quality, and enhance the delineation of organ boundaries. 没有加速收敛，但是鲁棒性很好，在不同模型上结果表现得都很好。
\item To fully exploit the location prior map, we design a crossing dual-encoder architecture consisting of four modules: a global encoder, a local encoder, a symmetric cross-attention module and a flow-based decoder. Ablation studies demonstrate that the proposed crossing dual-encoder design significantly enhances the segmentation performance of the model. Furthermore, we validate the effectiveness of the proposed symmetric cross-attention module. % 我们进一步对称的cross的作用。
\end{itemize}
% document is a template for \LaTeX.
% You are encouraged to use it to prepare your manuscript.
% If you are reading a paper or PDF version of this document, please download the 
% \LaTeX .zip file from the IEEE Web site at \underline
% {https://www.embs.org/tmi/authors-instructions/} to prepare your manuscript.
% You can also explore using the Overleaf editor at 
% \underline
% {https://www.overleaf.com/blog/278-how-to-use-overleaf-with-}\discretionary{}{}{}\underline
% {ieee-collabratec-your-quick-guide-to-getting-started\#.}\discretionary{}{}{}\underline{xsVp6tpPkrKM9}

% \section{Guidelines for Manuscript Preparation}
\section{Relation work}
\subsection{Multi-Organ Segmentation Models}
Multi-organ segmentation is challenging due to anatomical variability in organ shape, size, and position. Recent models based on CNNs, Transformers, and their hybrids have shown promising results. CNN-based methods\cite{alom2019recurrent,ronneberger2015unet,li2023sunet,zhou2018unetplusplus} excel at local feature extraction. For example, Alom et al.\cite{alom2019recurrent} proposed R2U-Net, which innovatively incorporates residual connections to alleviate the vanishing gradient problem in deep networks and enhance feature propagation. However, CNN-based models still face inherent limitations in modeling long-range dependencies, while transformers are effective for long-range context modeling. For example, Liu et al.\cite{liu2021swintransformer} proposed the Swin Transformer, which combines local window-based self-attention to significantly reduce computational complexity while preserving strong global modeling capabilities, and MISSFormer\cite{huang2022missformer} enhances local detail with multi-scale fusion. However, Transformer-based models often struggle to capture fine-grained local details, which are critical for precise boundary delineation in medical image segmentation. Hybrid models combine the strengths of CNNs and Transformers. TransUNet\cite{chen2021transunet} integrates CNN features with Transformer context. ATTransUNet\cite{li2023attransunet} further improves representation via adaptive attention. Swin-Unet\cite{cao2021swinunet} extends the Swin Transformer into a U-shaped architecture tailored for medical image segmentation, enabling hierarchical feature extraction with enhanced global-local contextual understanding.

\subsection{Architecture-Driven Segmentation}
% 第一个挑战的解决方案，医学图像低对比度，器官和器官之间分割的不准确。很多人尝试设计更复杂的网络架构。例如****。我们的dual encoder能够充分的考虑器官的信息。提高分割时边缘的准确性。
% 第一句虽然基于UNET展现出很高的表现，但是一些研究表明通过设计更加复杂的网络能够抓取更加复杂的语义特征以解***问题。
Although U-Net variants have demonstrated persuasive performance in medical image segmentation, recent studies \cite{rahman2024gcascade,sun2024paratranscnn,xie2021cotr} suggest that more sophisticated network architectures can better capture complex semantic features to improve the accuracy of segmentation. For example, Zhang et al.\cite{rahman2024gcascade} introduced G-CASCADE, which enhances the decoder with cascaded graph convolutional layers to improve the model’s boundary modeling capabilities. Sun et al.\cite{sun2024paratranscnn} introduced ParaTransCNN, a parallel CNN-Transformer architecture that adds an auxiliary encoder for capturing fine-grained details, achieving more accurate segmentation in complex anatomical regions with subtle boundary variations. Xie et al.\cite{xie2021cotr} proposed CoTr, a dual-encoder framework that integrates a CNN encoder with a deformable Transformer encoder. By introducing a DeTrans module, the model focuses attention on key spatial regions, effectively capturing global context for improved multi-organ segmentation. These studies collectively demonstrate that more sophisticated network architectures can effectively enhance model generalization and improve the accuracy of organ boundary segmentation. Despite architectural innovations, existing methods generally lack explicit modeling of organ spatial priors, limiting their ability to exploit the predictable anatomical layout of organs. To overcome this limitation, we propose a crossing dual-encoder architecture that incorporates a prior map generated in the first stage to guide the local encoder, thereby enhancing feature extraction and improving segmentation accuracy.

\subsection{Anatomical Prior-Guided Segmentation}
% 腹部医学图像和自然图像不同的是****，腹部医学影像的包含若干器官，这些器官具有解刨规律，具体参考引言中多器官分割的第二个难点。介绍别人的相关论文****。最后介绍我们的方法和别人方法的不通电。 
Compared to natural images, organs in medical images often follow predictable anatomical patterns, such as consistent shape and spatial arrangement. Consequently, some studies have explored the use of prior knowledge to capture the spatial distribution of target structures, aiming to improve segmentation accuracy. For example, Huang et al.\cite{huang2021atlas} proposed Deep Atlas Prior, which for the first time integrates anatomical priors into multi-organ segmentation via a dedicated prior-aware loss function. Their approach improves boundary precision and spatial consistency across organs. Wen et al.\cite{wen2024dclnet} introduced DCL‑Net, a dual-stage contrastive learning framework where the first stage generates a coarse mask that serves as a spatial prior to guide the second stage’s organ-aware feature refinement. This design significantly enhances structure-specific feature learning and improves segmentation accuracy, particularly in low-contrast regions. Zhong et al.\cite{zhong2025pgsam} proposed PG-SAM, a framework that integrates fine-grained semantic priors extracted by a medical-domain large language model to enhance the Segment Anything Model (SAM) for multi-organ segmentation. PG-SAM demonstrates strong generalization across unseen anatomical structures and imaging modalities, achieving state-of-the-art performance on multiple benchmarks. Jiang et al.\cite{jiang2020twostage} proposed a two-level cascading U-Net, which uses a U-Net variant for the coarse segmentation in the first stage, then concatenates the result with the original input to utilize automatic context for fine-tuning in the second stage. These studies highlight the promising potential of incorporating prior knowledge into multi-organ segmentation frameworks. Inspired by these advancements, we propose an SP-Net, which introduces a Plug-and-Play pretrained segmentation network to produce the interest location priors. 
% Do not change the template font sizes or line spacing to squeeze more text into a limited number of pages.
% The preferred font is 10-pt Times New Roman. Use italics for emphasis; do not underline words.

\section{Method}

\begin{figure*}[!t]
    \centering
    \includegraphics[width=0.95\textwidth]{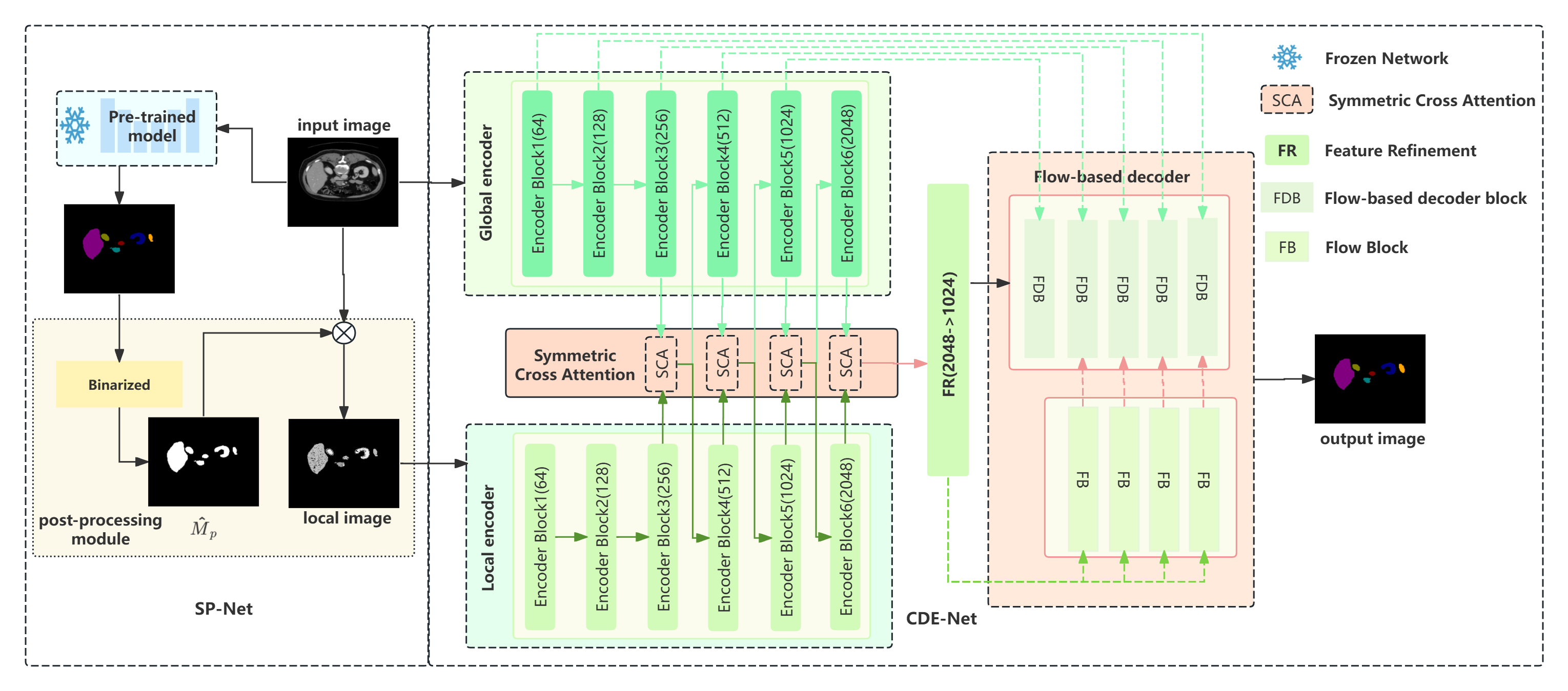}
    \caption{Overview of the SPG-CDENet}
    \label{fig:2}
\end{figure*}
Given a grayscale medical image $x \in \mathbb{R}^{H \times W \times 1}$, where $H$ and $W$ denote the image height and width, our proposed SPG-CDENet enhances multi-organ segmentation by explicitly incorporating prior anatomical cues. As shown in Fig.~\ref{fig:2}, SPG-CDENet consists of two components: a Spatial Prior Network (SP-Net) and a Crossing Dual-Encoder Network (CDE-Net). SP-Net takes $x$ as input and generates a coarse spatial prior $\hat{M}_{p}$ to highlight potential the ROI. CDE-Net then uses both $x$ and $\hat{M}_{p}$ to predict the final segmentation mask. CDE-Net includes two parallel encoders: a global encoder for capturing semantic features from the entire image and a local encoder focused on region-specific features guided by $\hat{M}_{p}$. To facilitate feature interaction between the two branches, we introduce a symmetric cross-attention mechanism, enabling effective fusion of global context and local details. The following sections detail the architecture of SPG-CDENet, including SP-Net, CDE-Net, and the proposed symmetric cross-attention strategy.

\subsection{Spatial Prior Network}

Multi-organ segmentation remains a challenging task due to the complex anatomical structures of the human body. Organs exhibit significant variability in size and shape across different organs. Furthermore, the low contrast between adjacent anatomical structures—including both organs and surrounding tissues—introduces ambiguity, especially in the segmentation of small or poorly defined organs. These factors often result in inaccurate localization and blurred boundaries in conventional segmentation models. To overcome these limitations, we propose SP-Net, a novel approach that generates a coarse segmentation mask to represent prior spatial information of the ROI. As shown in Fig.~\ref{fig:2}, SP-Net comprises two main components: a pre-trained multi-organ segmentation model and a post-processing module. The segmentation model takes the entire image $x$ as input and generates a multi-class segmentation mask, where each label corresponds to a specific organ. The post-processing module then binarizes this mask to distinguish the ROI from the background. The resulting binary mask serves as a spatial prior for the subsequent segmentation stage, guiding the model to focus on anatomically relevant regions and mitigating interference caused by inter-class ambiguity. This process can be formulated as follows:\begin{equation}
    x_{l} = x \times \mathbb{B}(PTM(x) > \tau)
\end{equation}where $\tau$ is a fixed threshold set to 0. $PTM(\cdot)$ denotes the pre-trained segmentation model, and $\mathbb{B}(\cdot)$ represents the binarization operation. $x_{l}$ is the output of the SP-Net, which provides spatial guidance for the downstream segmentation model.

\subsection{Crossing Dual Encoder-based Segmentor}

% \begin{figure*}[t!]
%     \centering
%     \includegraphics[width=1\textwidth]{LaTeX/block.png}
%     \caption{The various block structure in our network}
%     \label{fig:3}
% \end{figure*}

% 介绍整体的架构，
% 我们的网络架构图三所示，介绍图三中每个大模块的功能。
As illustrated in the Fig.~\ref{fig:2}, the proposed CDE-Net takes the input and local images as input and try to predict a mask about different organ. Proposed consists of three parts: dual encoder network, a flow-based decoder, and a symmetric crossing attention module. To be specific, we device a dual-encoder, the global encoder takes the original grayscale image as input, while the local encoder processes local image generated in the SP-Net. To strengthen the feature interaction between the two encoders, we proposed the symmetric cross-attention mechanism to integrate the features between the two encoders. For better global semantic guidance during decoding, we proposed the flow-based decoder, which propagate the final-layer features from the encoder to each layer of the decoder. This design allows the decoder to access high-level contextual information at all stages, thereby improving the segmentation of small or ambiguous structures. In the following, we provide a detailed description of the three key components of our CDE-Net.

% This section provides a detailed overview of the proposed CDE-Net architecture, as illustrated in the Fig.~\ref{fig:3}. In our dual-encoder design, the global encoder takes the original grayscale image as input, while the local encoder processes \(\hat{M}_{p}\) generated in the SP-Net. Both encoders employ ResNet-50 to extract hierarchical features. We begin by introducing the structure of dual encoder. Then, we describe the cross attention mechanism for effective feature fusion and the flow-based decoder that integrates multi-source semantic information across scales.

\subsubsection{Dual Encoder Network}

As shown in Fig.~\ref{fig:2}, we employ two parallel networks as encoders: one processes the entire image to extract global semantic features, while the other focuses on the local image to extract fine-grained local features. Each encoder follows the layered ResNet-50 structure, producing hierarchical features with progressively decreasing spatial resolution and increasing channel dimensions. The two encoders share an identical architecture. Taking the global encoder as an example, the output of each layer can be represented as : 
\begin{equation}
    F_{Enc}^{GE(i)}=\left\{\begin{array}{ll}
\operatorname{Res}_{i}(x), & i=0 \\
\operatorname{Res}_{i}\left(F_{Enc}^{GE(i-1)}\right), & i=1,2,3,4,5 
\end{array}\right.
\end{equation}
% \operatorname{Bridge}\left(F_{E n c}^{5}\right), & i=5
Where \({Res}_{i}(\cdot)\) represents the i-th layer of Resnet-50 Block. The local encoder shares the same architecture as the global encoder, with the only difference being that it takes the local image as input to the first layer.
The feature maps from the final layers of the dual encoders are concatenated and then passed through a feature refinement (FR) module, which performs feature compression and channel adjustment. This process can be formalized as:
\begin{equation}
    F_{global}=FR(Cat[F_{Enc}^{GE(5)},F_{Enc}^{LE(5)}])
\end{equation}
FR module consists of two consecutive Conv Blocks with identical structures, as illustrated in Fig.~\ref{fig:3}. Each Conv Block is composed of a $3\times3$ convolution followed by a BatchNorm2d layer and a ReLU activation. We denote the output of the FR module as $F_{global} \in \mathbb{R}^{H \times W \times C}$, which serves as the high-level semantic representation for the decoder.

\subsubsection{Symmetric Cross-Attention Module}
\begin{figure}[t!]
    \centering
    \includegraphics[width=0.45\textwidth]{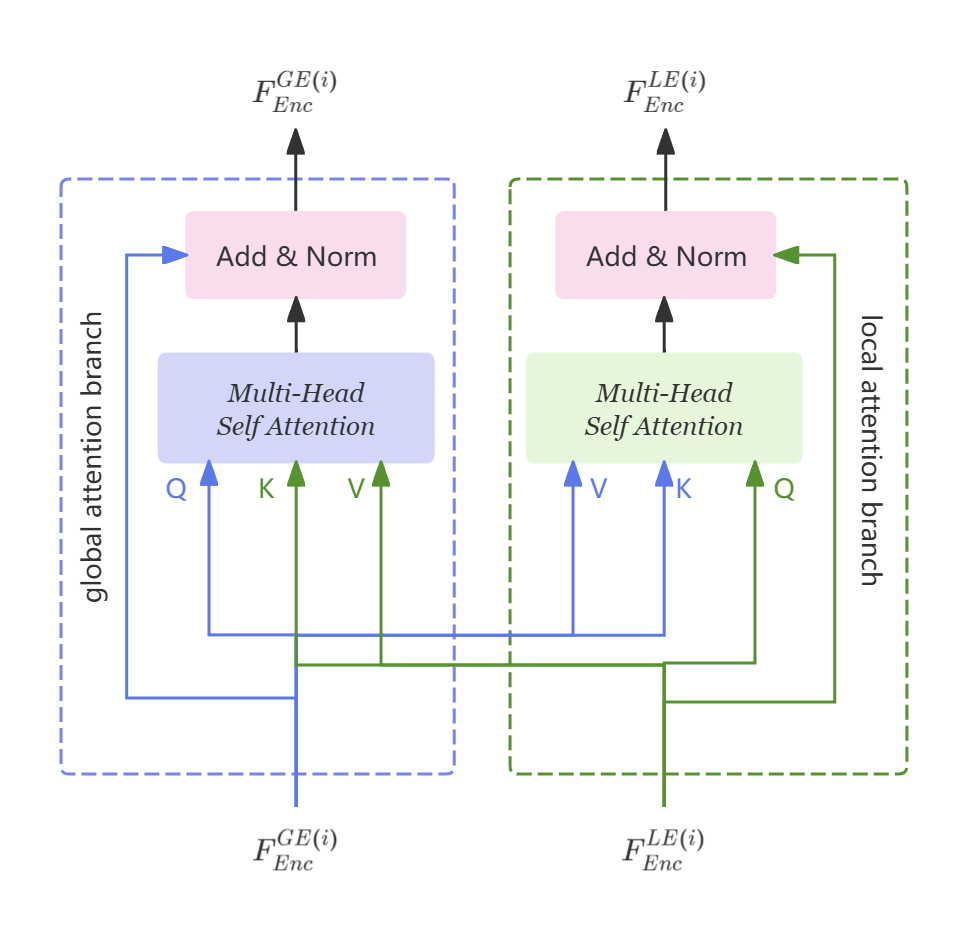}
    \caption{Symmetric Cross-Attention Module}
    \label{fig:3}
\end{figure}

% We only apply cross-attention in these final four layers for two key reasons: first, utilizing deeper semantic features helps reduce computational cost; second, to progressively leverage semantic features at different levels.
As shown in Fig. \ref{fig:2}, we apply the Symmetric Cross-Attention (SCA) module between the third and sixth layers of the global and local encoders to facilitate effective communication between them. The architecture of proposed SCA module, depicted in Fig. \ref{fig:3}, contains two branches: global  attention branch and local attention branch. In the global attention branch, a multi-head attention is applied to allow the local encoder to be guided by global context. Specifically, semantic features from the global encoder are used as queries (Q), while the semantic features from the local encoder serve as keys (K) and values (V). The output of the multi-head attention is then combined with the original semantic features through a residual connection, yielding the final result. Taking the i-th layer as an example, this process can be expressed as: 
\begin{equation}
Q = F_{Enc}^{GE(i)} W^{QG}, K = F_{Enc}^{LE(i)} W^{KG}, V = F_{Enc}^{LE(i)} W^{VG},
\end{equation}
\begin{equation}
F_{Enc}^{GE(i+1)} = \mathrm{Softmax}\!\left(\frac{Q K^\top}{\sqrt{d}}\right) V + F_{Enc}^{GE(i)}
\end{equation}where $W^{QG}$, $W^{KG}$, $W^{VG}$ are learnable parameters in the multi-head attention mechanism of the global attention branch.

The local attention branch follows same architecture as the global attention branch. However, in the local attention, semantic features captured from the local encoder are used as queries (Q), while semantic features captured from the global encoder serve as keys (K) and values (V). This process can be formally expressed as:\begin{equation}
Q = F_{Enc}^{LE(i)} W^{QL},  K =  F_{Enc}^{GE(i)} W^{KL},  V = F_{Enc}^{GE(i)} W^{VL},
\end{equation}
\begin{equation}
F_{Enc}^{LE(i+1)} = \mathrm{Softmax}\!\left(\frac{Q K^\top}{\sqrt{d}}\right) V + F_{Enc}^{LE(i)}
\end{equation}where $W^{QL}$, $W^{KL}$, $W^{VL}$ are learnable parameters in local-guided attention branch.

\subsubsection{Flow-Based Decoder}
\begin{figure}[t!]
    \centering
    \includegraphics[width=0.5\textwidth]{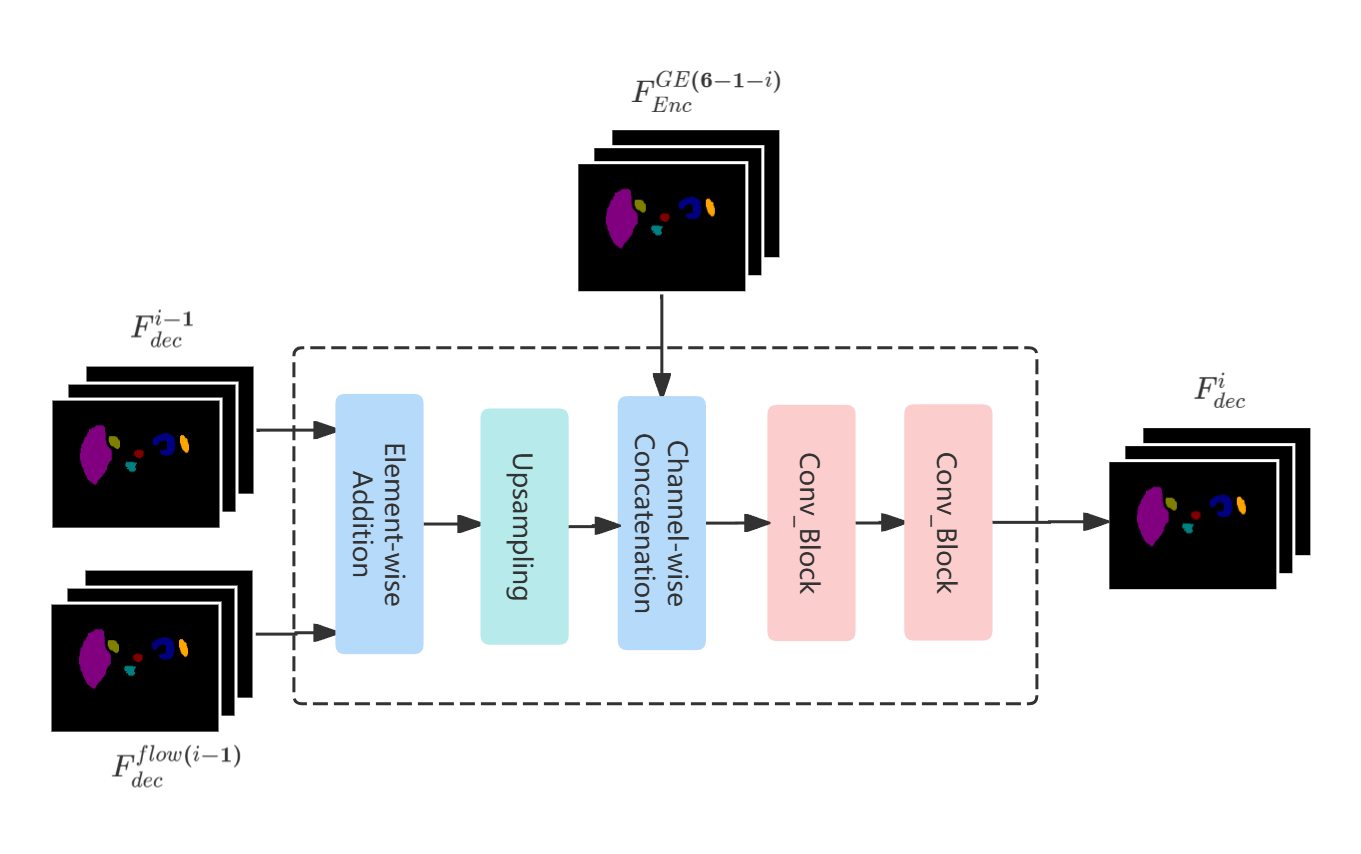}
    \caption{Overview of the Flow-based Decoder. The decoder takes three inputs — semantic features from the previous decoder layer $F^{i-1}_{Dec}$, the encoder $F^{GE(6-1-i)}_{Enc}$ , and the Flow Block layer $F^{flow(i-1)}_{Dec}$, respectively.}
    \label{fig:4}
\end{figure}
\begin{table*}[tp]
\centering
\caption{Comparison on the Synapse multi-organ CT dataset (average DSC score \% and average Hausdorff distance in mm, with higher DSC and lower HD being preferred, and the Spatial Prior Network of SPG-CDENet (Ours) is U-Net)}
\label{tab:1}
\begin{tabular}{l|c|c|cccccccc}
\hline
\textbf{Methods} & \textbf{DSC(\%)}↑ & \textbf{HD↓} & \textbf{Aorta} & \textbf{Gallbladder} & \textbf{Kidney(L)} & \textbf{Kidney(R)} & \textbf{Liver} & \textbf{Pancreas} & \textbf{Spleen} & \textbf{Stomach} \\
\hline
V-Net \cite{milletari2016vnet} & 68.81 & - & 75.34 & 51.87 & 77.10 & 80.75 & 87.84 & 40.05 & 80.56 & 56.98 \\
DARR \cite{alom2019recurrent} & 69.77 & - & 74.74 & 53.77 & 72.31 & 73.24 & 94.08 & 54.18 & 89.90 & 45.96 \\
R50 U-Net \cite{chen2021transunet} & 74.68 & 36.87 & 87.47 & 66.36 & 80.60 & 78.19 & 93.74 & 56.90 & 85.87 & 74.16 \\
U-Net \cite{ronneberger2015unet} & 76.85 & 39.70 & 89.07 & 69.72 & 77.77 & 68.60 & 93.43 & 53.98 & 86.67 & 75.58 \\
R50 Att-UNet \cite{chen2021transunet} & 75.57 & 36.97 & 55.92 & 63.91 & 79.20 & 72.71 & 93.56 & 49.37 & 87.19 & 74.95 \\
Att-UNet \cite{oktay2018attentionunet} & 77.77 & 36.02 & 89.55 & 68.88 & 77.98 & 71.11 & 93.57 & 58.04 & 87.30 & 75.75 \\
R50 ViT \cite{chen2021transunet} & 71.29 & 32.87 & 73.73 & 55.13 & 75.80 & 72.20 & 91.51 & 45.99 & 81.99 & 73.95 \\
TransUNet \cite{chen2021transunet} & 77.48 & 31.69 & 87.23 & 63.13 & 81.87 & 77.02 & 94.08 & 55.86 & 85.08 & 75.62 \\
TransNorm \cite{azad2022transnorm} & 78.40 & 30.25 & 86.23 & 65.10 & 82.18 & 78.63 & 94.22 & 55.34 & 89.50 & 76.01 \\
Swin U-Net \cite{cao2021swinunet} & 79.13 & 21.55 & 85.47 & 66.53 & 83.28 & 79.61 & 94.29 & 56.58 & 90.66 & 76.60 \\
TransDeepLab \cite{azad2022transdeeplab} & 80.16 & 21.25 & 86.04 & 69.16 & 84.08 & 79.88 & 93.53 & 61.19 & 89.00 & 78.40 \\
HiFormer \cite{heidari2023Hiformer} & 80.39 & 14.70 & 86.21 & 65.69 & 85.23 & 79.77 & 94.61 & 59.52 & 90.99 & 81.08 \\
MISSFormer \cite{huang2022missformer} & 81.96 & 18.20 & 86.99 & 68.65 & 85.21 & 82.00 & 94.41 & 65.67 & 91.92 & 80.81 \\
TransCeption \cite{azad2023transception} & 82.24 & 20.89 & 87.60 & 71.82 & 86.23 & 80.29 & 95.01 & 65.27 & 91.68 & 80.02 \\
DAE-Former \cite{azad2023daeformer} & 82.63 & 16.39 & 87.84 & 71.65 & 87.66 & 82.39 & \textcolor{blue}{95.08} & 63.93 & 91.82 & 80.77 \\
FCT \cite{tragakis2023fullyconvolutional} & 83.53 & - & \textbf{89.85} & 72.73 & \textcolor{blue}{88.45} & \textbf{86.60} & \textbf{95.62} & 66.25 & 89.77 & 79.42 \\
ParaTransCNN \cite{sun2024paratranscnn} & 83.86 & 15.86 & 88.12 & 68.97 & 87.99 & 83.84 & 95.01 & 69.79 & \textbf{92.71} & 84.43 \\
AHGNN \cite{chai2024hypergraph} & \textcolor{blue}{84.03} & \textcolor{blue}{13.26} & \textcolor{blue}{89.27} & \textcolor{blue}{74.53} & 86.99 & 83.49 & 95.03 & \textcolor{blue}{69.89} & 92.38 & \textcolor{blue}{84.86} \\
\textbf{SPG-CDENet (Ours)} & \textbf{85.97} & \textbf{12.75} & 87.03 & \textbf{75.36} & \textbf{89.77} & \textcolor{blue}{86.03} & 94.72 & \textbf{72.78} & \textcolor{blue}{92.69} & \textbf{89.40} \\
\hline
\end{tabular}
\end{table*}

To enhance the global semantic guidance during decoding, we design a flow-based decoder as shown in the Fig.~\ref{fig:2}, which comprises a backbone decoder and a flow module. The flow module delivers multi-scale high-level semantic features from $F_{global}$ to each decoding layer, with a upsampling layer and two consecutive Conv Blocks, thereby reinforcing semantic consistency during reconstruction. The detailed implementation of flow module is described as follows:\begin{equation}
\{F^{flow(i)}_{Dec}\}_{i=1}^4 = FB_{s_i}(F_{global}, c_i)
\end{equation}where $FB_{s_i}(\cdot)$ represents the flow module at the i-th level. $c_i$ and $s_i$ represent the number of output channels and the upsampling factor at that level, respectively. Specifically, the configuration is defined as \((c_i, s_i) \in \{(1024,2), (512,4), (256,8), (128,16)\}\). 
% The first decoder layer takes two inputs: the global feature $F_{global}$ and the skip connection feature $F_{GE}^4$: denotes bilinear upsampling with factor $s$, $Conv(\cdot)$  input and output channels set as $C_0 \times2^{4-i}$, and $i \in \{1,2,3,4\}$ corresponds to upsampling factors $\{2, 4, 8, 16\}$

% \begin{equation}
% \{F_i\}_{i=1}^4 = {Up}_{s_i}\big(\mathrm{Flow}(F_{global}, c_i)\big) 
% \end{equation}

After generating the multi-scale flow features, we incorporate them into the decoding pathway to construct the complete flow-based decoder. Except for the first layer, as shown in the Fig.\ref{fig:4}, each decoding layer $F^{i}_{Dec}$ integrates three types of input features: the output from the previous decoding layer $F^{i-1}_{Dec}$, the skip connection feature from the global encoder $F^{GE(6-1-i)}_{Enc}$, and the flow feature at the corresponding scale $F^{flow(i-1)}_{Dec}$. The overall decoding process can be formulated as follows:
\begin{equation}
\resizebox{\linewidth}{!}{$
F^{i}_{Dec} = 
\left\{
\begin{array}{ll}
\operatorname{FDB}(F_{global}, F^{GE(6-1-i)}_{Enc}), & i=1 \\[6pt]
\operatorname{FDB}(F^{i-1}_{Dec}, F^{GE(6-1-i)}_{Enc} + F^{flow(i-1)}_{Dec}), & i=2,3,4,5
\end{array}
\right.
$}
\end{equation}where $FDB(\cdot)$ represents flow-based decoder block. Semantic features $F^{5}_{Dec}$ is fed into an output layer to generate the final segmentation mask. The output layer consists of two Conv Blocks followed by a sigmoid activation function. This process can be formulated as \begin{equation}
    \hat{M} = \sigma(OL(F^{5}_{Dec}))),
\end{equation}where $OL$ is the output layer. $\hat{M}$ denotes the final predicted mask generated by our model.

% $C=6$ denotes the total number of encoder layers and $\hat{M}=F^{5}_{Dec}$ denotes the final predicted segmentation mask.  Here, $c_i$ denotes the output channel dimension of the i-th IPPM branch, which progressively reduces from 1024 to 128, $s_i$ indicates the upsampling factor applied to the branch output, with values of {2,4,8,16} corresponding to different spatial scales.

\subsection{Loss Function}
For training our SPG-CDENet network, we employ a combination of Dice loss and Cross Entropy as the loss function, defined as follows:
\begin{equation}
\mathcal{L} = \lambda_1 \mathcal{L}_{\text{dice}}(M,\hat{M}) + \lambda_2 \mathcal{L}_{\text{ce}}(M,\hat{M})
\end{equation}here \(\mathcal{L}\) is the total loss, with \(\lambda_1\) and \(\lambda_2\) as the weight coefficients, with respective set to 0.4 and 0.6 based on experimental results. And $M$ represent the ground-truth segmentation mask provided by expert annotations.
\section{Experimental}
\subsection{Experimental Settings }

\subsubsection{Dataset}
We evaluate our proposed method on two widely-used benchmarks, the Synapse multi-organ CT dataset from the MICCAI 2015 Multi-Atlas Labeling Beyond the Cranial Vault Challenge (BTCV) and the ACDC cardiac MRI dataset\cite{bernard2018deep}, followed by a detailed description of their data splits and evaluation protocols. \textbf{The Synapse multi-organ segmentation dataset} comprises 30 abdominal CT scans with a total of 3,779 axial slices. Following the widely adopted experimental protocol, we utilize 18 cases for training and 12 cases for testing, consistenting with the previous method \cite{cao2021swinunet,huang2022missformer,chen2021transunet}. The evaluation is conducted on eight abdominal organs, including the aorta, gallbladder, spleen, left kidney, right kidney, liver, pancreas, and stomach. \textbf{The ACDC dataset} consists of 100 short-axis cardiac MR images collected from different patients under breath-hold conditions. Each image is manually annotated for three cardiac structures: the left ventricle (LV), right ventricle (RV), and myocardium (MYO). In line with previous works \cite{cao2021swinunet,huang2022missformer,chen2021transunet}, we use 70 images for training, 10 for validation, and 20 for testing.

\subsubsection{Evaluation Metrics}
To quantitatively assess the segmentation performance of our method, we adopt the average Dice Similarity Coefficient (DSC) and the average Hausdorff Distance (HD) as evaluation metrics. The DSC measures the overlap ratio between the predicted segmentation and the ground truth, while the HD evaluates the boundary agreement between them. Higher DSC and lower HD indicate better segmentation accuracy and boundary consistency.
% The average Dice Similarity Coefficient (DSC) is adopted as the evaluation metric to assess the segmentation performance of our method, 

\subsubsection{Implement Details}
All experiments were conducted on a single NVIDIA RTX 4090 GPU, with network parameters optimized using the Stochastic Gradient Descent (SGD) optimizer. The input image and local image size was set to $224 \times 224$, the initial learning rate is 5e-3, the momentum 0.9, and the weight decay 1e-4. In all experiments, we applied basic data augmentation techniques, including random rotations and flips. It is worth noting that, we apply consistent data augmentation to both global and local images for enhancing the model’s robustness to local image variations. 

\subsubsection{Comparison of experimental models}
We compare SPG-CDENet with CNN-based segmentation networks like R50 U-Net\cite{chen2021transunet}, U-Net\cite{ronneberger2015unet}, Att-UNet\cite{oktay2018attentionunet}, pure Transformer segmentation network systems such as Swin U-Net\cite{cao2021swinunet}, TransDeepLap\cite{azad2022transdeeplab}, MISSFormer\cite{huang2022missformer}, DAE-Former\cite{azad2023daeformer}, the mixed CNN and Transformer models like TransUNet\cite{chen2021transunet}, HiFormer\cite{heidari2023Hiformer}, FCT\cite{tragakis2023fullyconvolutional}, TransCeption\cite{azad2023transception}, and dual encoder models like ParaTransCNN\cite{sun2024paratranscnn}.

\begin{figure*}[tp!]
    \centering
    \includegraphics[width=1\textwidth]{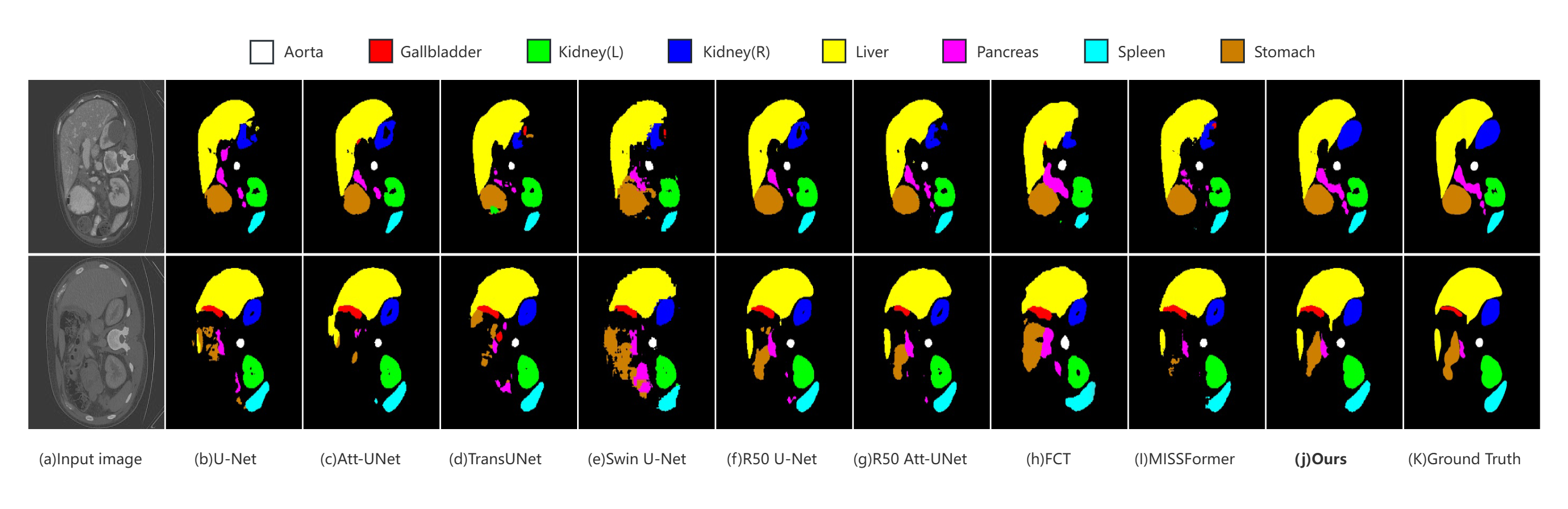}
    \caption{Qualitative results of different models on the Synapse dataset, from left to right: Input image; U-Net\cite{ronneberger2015unet}; Att-UNet\cite{oktay2018attentionunet}; TransUNet\cite{chen2021transunet}; Swin U-Net\cite{cao2021swinunet}; R50 U-Net\cite{chen2021transunet}; R50 Att-UNet\cite{chen2021transunet}; FCT\cite{tragakis2023fullyconvolutional}; MISSFormer\cite{huang2022missformer}; SPG-CDENet (Ours); Ground Truth.}
    \label{fig:5}
\end{figure*}

\begin{figure*}[tp!]
    \centering
    \includegraphics[width=1\textwidth]{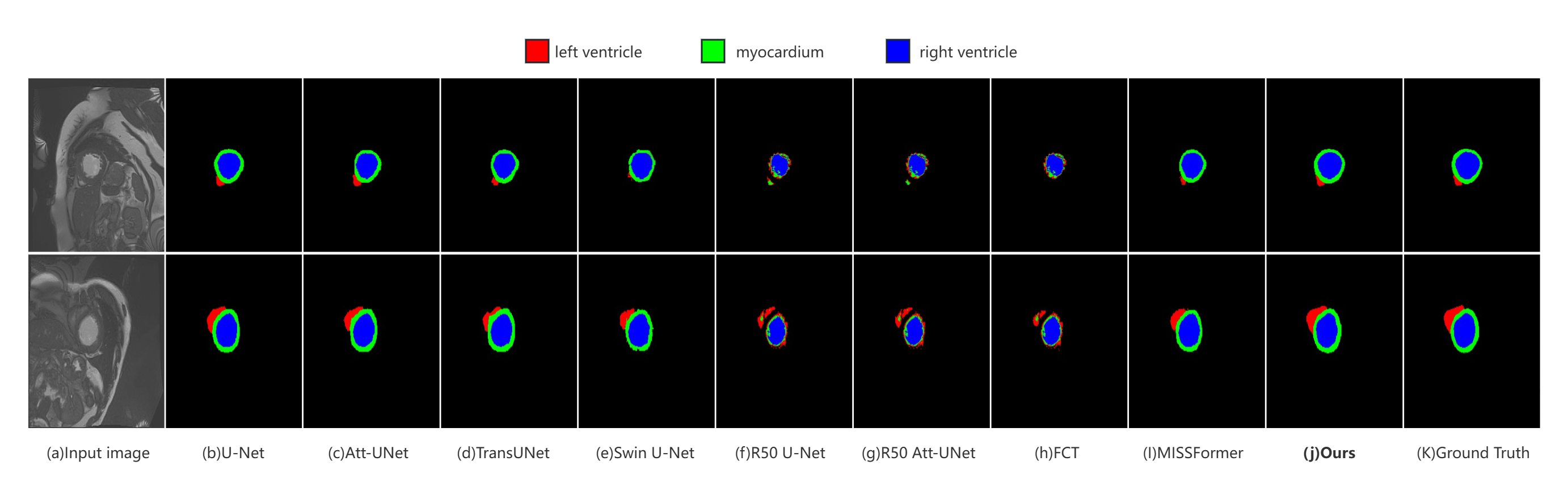}
    \caption{Qualitative results of different models on the  ACDC dataset, from left to right: Input image; U-Net\cite{ronneberger2015unet}; Att-UNet\cite{oktay2018attentionunet}; TransUNet\cite{chen2021transunet}; Swin U-Net\cite{cao2021swinunet}; R50 U-Net\cite{chen2021transunet}; R50 Att-UNet\cite{chen2021transunet}; FCT\cite{tragakis2023fullyconvolutional}; MISSFormer\cite{huang2022missformer}; SPG-CDENet (Ours); Ground Truth.}
    \label{fig:6}
\end{figure*}

\subsection{Comparison with state-of-the-art methods}
We compare SPG-CDENet with the state-of-the-art models on the Synapse dataset and ACDC dataset, quantitative results are shown in Fig. ~\ref{fig:5} and Fig. ~\ref{fig:6}, while qualitative results are presented in Tables 1 and 2.

\subsubsection{Results on Synapse Multi-Organ Segmentation}

In Table 1, our SPG-CDENet uses UNet as the SP-Net. Our SPG-CDENet achieves the best overall performance, attaining a DSC of 85.97\%, which surpasses the second-best method by 1.94\%, and ranking first in terms of the HD with a score of 12.75, which surpasses the second-best method by 0.51. This demonstrates the effectiveness of our approach. Our model demonstrates significant improvements in organs such as the gallbladder, kidney(L), pancreas, and stomach, with increases of 0.83\%, 1.32\%, 2.89\%, and 4.54\%, respectively, compared to the second-best model. Additionally, the visualization of segmentation results is shown in Fig.~\ref{fig:5}. Our model achieves superior overall performance and more precise local segmentation on the Synapse dataset. In particular, it produces accurate and complete delineations of small organs, such as the pancreas and stomach, which further confirms the consistency between qualitative and quantitative assessments. Furthermore, compared to the U-Net model, our approach achieves improvements in every organ except for the Aorta, which enable the model to capture both organ-specific positional priors and image-driven appearance features. This result demonstrates the effectiveness of our method by integrating SP-Net with a cross dual-encoder architecture.

% As illustrated in Fig.~\ref{fig:5}, our model achieves the best overall segmentation performance, with smooth organ boundaries. It demonstrates accurate and complete segmentation of small organs, such as the pancreas and stomach. while also accurately segmenting other organs. Meanwhile, we have no classification errors.
% Our method ranks second in the segmentation of the right kidney and spleen, while maintaining consistently strong performance across the remaining organs. This demonstrates that SPG-CDENet excels not only in handling small organs but also in adapting to the complex segmentation environment of multiple organs. What's more, compared to the U-Net model, our approach achieves improvements in every organ except for the Aorta, which enable the model to capture both organ-specific positional priors and image-driven appearance features.

% The model effectively manages morphological variations across different organs, ensuring stable and reliable segmentation results.

\begin{table}[tp]
\centering
\caption{Comparison with the state-of-the-art methods on the ACDC dataset (average DSC score \% , with higher DSC being preferred, and the Spatial Prior Network of SPG-CDENet (Ours) is U-Net)}
\label{tab:2}
\begin{tabular}{l|c|ccc}
\hline
\textbf{Methods} & \textbf{DSC(\%)}↑ & \textbf{RV} & \textbf{Myo} & \textbf{LV}  \\
\hline
R50-U-Net \cite{chen2021transunet} & 87.55 & 87.10 & 80.63 & 94.92  \\
U-Net \cite{ronneberger2015unet} & 87.79 & 85.48 & 82.90 & 94.97  \\
R50-AttnUNet \cite{chen2021transunet} & 86.75 & 87.58 & 79.20 & 93.47  \\
Att-UNet \cite{oktay2018attentionunet} & 89.58 & 87.89 & 85.56 & 95.29  \\
TransUNet \cite{chen2021transunet} & 89.71 & 88.86 & 84.53 & 95.73  \\
Swin U-Net \cite{cao2021swinunet} & 90.00 & 88.55 & 85.62 & 95.83 \\
% TransDeepLab \cite{azad2022transdeeplab} & 90.41 & 91.45 & 84.54 & 95.23  \\
% HiFormer \cite{heidari2023hiformer} & 90.12 & 91.06 & 84.54 & 94.77  \\
MISSFormer \cite{huang2022missformer} & 90.86 & 89.55 & 88.04 & 94.99  \\
% TransCeption \cite{azad2023transception} & 88.47 & 87.88 & 82.87 & 94.66  \\
% DAE-Former \cite{azad2022daeformer} & 89.78 & 89.91 & 84.38 & 95.04  \\
ParaTransCNN \cite{sun2024paratranscnn}  & 91.31 & \textbf{92.76} & 85.84 & 95.34  \\
AHGNN \cite{chai2024hypergraph} & 92.02 & - & - & - \\
FCT \cite{tragakis2023fullyconvolutional} &  \textcolor{blue}{92.84} &  92.02 & \textcolor{blue}{90.61} & \textcolor{blue}{95.89}  \\
\textbf{SPG-CDENet (Ours)} & \textbf{94.25} & \textcolor{blue}{92.61} & \textbf{91.74} & \textbf{98.40}  \\
\hline
\end{tabular}
\end{table}

\subsubsection{Results on ACDC Segmentation}
We also evaluated our method on the ACDC dataset, and as shown in the Table \ref{tab:2}, SPG-CDENet achieves the highest segmentation accuracy with a DSC of 94.25\%, surpassing the second-best model by 1.41\%. Compared with the U-Net (SP-Net), our model achieves consistent improvements across all cardiac substructures on the ACDC dataset, demonstrating the overall effectiveness and strong generalization ability of the proposed SPG-CDENet framework. The visual results, as shown in Fig.~\ref{fig:6}, demonstrate that SPG-CDENet produces the smoothest boundaries and the most complete segmentations, highlighting the model's strong generalizability and robustness.

\subsection{Ablation Studies On Components}

To investigate the impact of the symmetric cross-attention module, the dual-encoder module and the spatial prior network on the performance of our model, we conducted structural ablation experiments on Synapse dataset and ACDC dataset. Based on the full SPG-CDENet model, We designed three ablation variants: Model-1 serves as the baseline, consisting of a ResNet-50 encoder combined with a flow-based decoder. Model-2 incorporates the spatial prior network and a local encoder dedicated to processing the prior map, where the features from the two encoders remain independent without interaction. Finally, SPG-CDENet enhances this design by integrating a symmetric cross-attention module to fuse the features extracted from the two encoders at each layer.

\begin{table*}[htbp]
\centering
\caption{Ablation Studies on Components(Synapse). SP-Net indicates the Spatial Prior Network, implemented using U-Net. SCA indicates the Symmetric Cross-Attention module. \checkmark indicates the component is included.}
\label{tab:3}
\resizebox{\textwidth}{!}{\begin{tabular}{c|c|c|cc|cccccccc}
\hline
\textbf{Model} & \textbf{SP-Net}  & \textbf{SCA} & \textbf{DSC(\%)}↑ & \textbf{HD↓} & \textbf{Aorta} & \textbf{Gallbladder} & \textbf{Kidney(L)} & \textbf{Kidney(R)} & \textbf{Liver} & \textbf{Pancreas} & \textbf{Spleen} & \textbf{Stomach} \\
\hline
Model-1        &                  &             & 80.09  & 21.11 & 85.74  &  58.73 & 81.53  &  77.36 & 94.16  & 64.07  & 92.55  & 86.56  \\
Model-2        & \checkmark       &             & 82.47  & 21.80 & 81.59 & 76.90 & 83.14 & 79.94 & 94.20 & 65.04 & 92.22 & 86.74 \\
SPG-CDENet        & \checkmark       & \checkmark  & \textbf{85.97} & \textbf{12.75} & \textbf{87.03} & \textbf{75.36} & \textbf{89.77} & \textbf{86.03} & \textbf{94.72} & \textbf{72.78} & \textbf{92.69} & \textbf{89.40} \\
\hline
\end{tabular}}
\end{table*}

\begin{table*}[!t]
\centering
\caption{Fusion Type Ablation Study On Synapse (The Spatial Prior Network is U-Net)}
\label{tab:5}
\begin{tabular}{c|c|c|cccccccc}
\hline
\textbf{Fusion Type} & \textbf{DSC(\%)}↑ & \textbf{HD↓} & \textbf{Aorta} & \textbf{Gallbladder} & \textbf{Kidney(L)} & \textbf{Kidney(R)} & \textbf{Liver} & \textbf{Pancreas} & \textbf{Spleen} & \textbf{Stomach} \\
\hline
\textbf{None}   & 82.47  & 21.80 & 81.59 & 76.90 & 83.14 & 79.94 & 94.20 & 65.04 & 92.22 & 86.74 \\
\textbf{Concat}   & 82.92 & 27.78 & 80.11 & 77.38 & 79.84 & 79.63 & 93.59 & 75.03 & 90.57 & 87.18 \\
\textbf{Cross-attention}  & 83.17 & 18.39 & 86.34 & 72.88 & 85.62 & 79.33 & 95.06 & 69.47 & 92.39 & 84.26 \\
\textbf{SPG-CDENet} & \textbf{85.97} & \textbf{12.75} & \textbf{87.03} & \textbf{75.36} & \textbf{89.77} & \textbf{86.03} & \textbf{94.72} & \textbf{72.78} & \textbf{92.69} & \textbf{89.40} \\
\hline
\end{tabular}
\end{table*}

\begin{table}[tp]
\centering
\caption{Ablation Studies on Components(ACDC). SP-Net indicates the Spatial Prior Network, implemented using U-Net. SCA indicates the Symmetric Cross-Attention module. \checkmark indicates the component is included.}
\label{tab:4}
\resizebox{0.45\textwidth}{!}{\begin{tabular}{c|c|c|c|ccc}
\hline
\textbf{Model} & \textbf{SP-Net}  & \textbf{SCA} & \textbf{DSC(\%)}↑ & \textbf{RV} & \textbf{Myo} & \textbf{LV} \\
\hline
Model-1        &                  &             & 90.37  & 89.02 & 85.51  & 96.58    \\
Model-2        & \checkmark       &             & 91.04  & 90.01 & 86.65 & 96.47 \\
SPG-CDENet        & \checkmark       & \checkmark  & \textbf{94.25} & \textbf{92.61} & \textbf{91.74} & \textbf{98.40} \\
\hline
\end{tabular}}
\end{table}

As presented in Table \ref{tab:3} and Table \ref{tab:4}, on both the Synapse and ACDC datasets, Model-1 achieved DSC scores of 80.09\% and 90.37\%, surpassing the performance of most CNN-based models. Adding the SP-Net and Local Encoder substantially improves the segmentation performance, confirming the effectiveness of anatomical prior guidance. Further integrating the two encoders through cross-attention fusion enables the full SPG-CDENet to fully exploit complementary features, yielding an additional 3.5\% DSC improvement and a 9.05 HD reduction on the Synapse dataset, along with consistent gains on the ACDC dataset. Notably, while SP-Net and the Local Encoder provide limited performance boost when used independently, but combining them with the SCA module results in a substantial performance improvement. This demonstrates the critical role of SCA module in fusing both global and local features effectively.
% which confirms that the IPPM enhances the information flow between high-level semantics and low-level structural details through a bottom-up aggregation pathway, ultimately improving the overall segmentation performance of the model.

\subsection{Ablation Study On Fusion Type}

To explore how different interaction methods between the global and local encoders affect model performance, we conducted an ablation study on various fusion types. We evaluate four variants, including \textbf{None}, \textbf{Concat}, \textbf{Cross Attention\cite{vaswani2017Attention}}, and \textbf{SCA modual}. When the fusion type is set to \textbf{None}, the two encoder streams remain independent without any feature interaction. This configuration corresponds to Model-2 in Section IV.C (Ablation Studies on Components) of the experimental results. For the \textbf{Concat} variant, the feature maps are concatenated along the channel dimension and subsequently compressed by a 1×1 convolution to form the fused representation. Under the \textbf{Cross Attention} setting, two encoder uses cross-attention to fuse features from two inputs, enhancing one feature map by attending to the other. SPG-CDENet refers to our full model that utilizes the SCA module to facilitate interaction between the two encoder streams.

The results are shown in Table \ref{tab:5} and Table \ref{tab:6}. It can be observed that under the \textbf{None} setting, the two encoder streams remain completely independent, resulting in the lowest average DSC and a relatively high HD, indicating that the lack of feature interaction limits model performance. Both concatenation and cross-attention fusion improve the overall segmentation performance, with the latter achieving more consistent and significant gains across all organs. For this phenomenon, we hypothesize that the possible reason is that concatenating features from the two encoders introduces redundant and misaligned information due to their differing segmentation scopes. Without effective interaction modeling, this leads to unstable boundary predictions and consequently a higher HD. Our proposed Symmetric Cross-Attention achieves the best segmentation results in both DSC and HD. It improves the performance for every organ, and also outperforms other fusion types, demonstrating the effectiveness of Symmetric Cross-Attention. Moreover, it has demonstrated strong performance on both the Synapse and ACDC datasets, further proving the robustness of Symmetric Cross-Attention.

 % Concat shows advantages on Myo (87.57\%) and LV (96.89\%). After introducing the single attention mechanism, the model performance significantly improved, especially for larger organs like Aorta (from 81.59\% to 86.34\%) and Kidney(L) (from 83.14\% to 85.62\%). However, the segmentation accuracy for smaller organs like Gallbladder remained low (72.88\%), indicating that while single attention aids in segmenting larger organs, it struggles to capture the complex features of smaller ones. These results suggest that simple fusion strategies fail to fully leverage the spatial priors provided by the prior map. The HD corresponding to the single cross attention fusion is also relatively high, indicating that while unidirectional attention helps capture the overall organ region, it lacks the reverse constraint and fails to adequately focus on the boundary regions

\begin{table}[tp]
\centering
\caption{Fusion Type Ablation Study On ACDC (The Spatial Prior Network is U-Net)}
\label{tab:6}
\begin{tabular}{c|c|ccc}
\hline
\textbf{Fusion Type} & \textbf{DSC(\%)}↑ & \textbf{RV} & \textbf{Myo} & \textbf{LV} \\
\hline
\textbf{None}  & 91.04  & 90.01 & 86.65 & 96.47 \\
\textbf{Concat}  & 91.64 & 90.46 & 87.57 & 96.89  \\
\textbf{Cross-attention}  & 92.53 & 91.77 & 88.70 & 97.13  \\
\textbf{SPG-CDENet} & \textbf{94.25} & \textbf{92.61} & \textbf{91.74} & \textbf{98.40} \\
\hline
\end{tabular}
\end{table}

\begin{table*}[tp!]
\centering
\caption{Ablation Study On Spatial Prior Network(Synapse)}
\label{tab:7}
\begin{tabular}{l|c|c|cccccccc}
\hline
\textbf{Model} & \textbf{DSC(\%)}↑ & \textbf{HD↓} & \textbf{Aorta} & \textbf{Gallbladder} & \textbf{Kidney(L)} & \textbf{Kidney(R)} & \textbf{Liver} & \textbf{Pancreas} & \textbf{Spleen} & \textbf{Stomach} \\
\hline
U-Net \cite{ronneberger2015unet} & 76.85 & 39.70 & 89.07 & 69.72 & 77.77 & 68.60 & 93.43 & 53.98 & 86.67 & 75.58 \\
\textbf{our + U-Net} & 85.97 & 12.75 & 87.03 & 75.36 & 89.77 & 86.03 & 94.72 & 72.78 & 92.69 & 89.40 \\
 & \textcolor{red}{(↑9.12)} & \textcolor{red}{(↓26.95)} & & \textcolor{red}{(↑5.64)} & \textcolor{red}{(↑12.00)} & \textcolor{red}{(↑17.43)} & \textcolor{red}{(↑1.29)} & \textcolor{red}{(↑18.80)} & \textcolor{red}{(↑6.02)} & \textcolor{red}{(↑13.82)} \\
\hline
Att-UNet \cite{oktay2018attentionunet} & 77.77 & 36.02 & 89.55 & 68.88 & 77.98 & 71.11 & 93.57 & 58.04 & 87.30 & 75.75 \\
\textbf{our + AttU-Net} & 85.91 & 13.23 & 86.70 & 75.24 & 89.84 & 85.88 & 94.72 & 73.03 & 92.59 & 89.27 \\
 & \textcolor{red}{(↑8.14)} & \textcolor{red}{(↓22.79)} & & \textcolor{red}{(↑6.36)} & \textcolor{red}{(↑11.86)} & \textcolor{red}{(↑14.77)} & \textcolor{red}{(↑1.15)} & \textcolor{red}{(↑14.99)} & \textcolor{red}{(↑5.29)} & \textcolor{red}{(↑13.52)} \\
\hline
% AHGNN \cite{chai2024hypergraph} & 84.03 & 13.26 & 89.27 & 74.53 & 86.99 & 83.49 & 95.03 & 69.89 & 92.38 & 84.86
% \textbf{our + AHGNN} & 85.92 & 14.01 & 86.74 & 75.38 & 89.73 & 85.75 & 94.68 & 73.22 & 92.64 & 89.24 \\
% \hline
TransUNet \cite{chen2021transunet} & 77.48 & 31.69 & 87.23 & 63.13 & 81.87 & 77.02 & 94.08 & 55.86 & 85.08 & 75.62 \\
\textbf{our + TransUNet}           & 86.02 & 12.63 & 86.75 & 75.53 & 89.76 & 86.12 & 94.80 & 72.70 & 92.89 & 89.63 \\
& \textcolor{red}{(↑8.54)} & \textcolor{red}{(↓19.06)} & & \textcolor{red}{(↑12.40)} & \textcolor{red}{(↑7.89)} & \textcolor{red}{(↑9.10)} & \textcolor{red}{(↑0.72)} & \textcolor{red}{(↑16.84)} & \textcolor{red}{(↑7.81)} & \textcolor{red}{(↑14.01)} \\
\hline
\end{tabular}
\end{table*}
\begin{table}[tp!]
\centering
\caption{Ablation Study On Spatial Prior Network(ACDC)}
\label{tab:8}
\begin{tabular}{l|c|ccc}
\hline
\textbf{Methods} & \textbf{DSC(\%)}↑ & \textbf{RV} & \textbf{Myo} & \textbf{LV}  \\
\hline
U-Net \cite{ronneberger2015unet} & 87.79 & 85.48 & 82.90 & 94.97  \\
\textbf{our + U-Net} & \textbf{94.25} & \textcolor{blue}{92.61} & \textbf{91.74} & \textbf{98.40} \\
 & \textcolor{red}{(↑6.46)} &  \textcolor{red}{(↑7.13)} & \textcolor{red}{(↑8.84)} & \textcolor{red}{(↑3.43)}  \\
\hline
Att-UNet \cite{oktay2018attentionunet} & 89.58 & 87.89 & 85.56 & 95.29  \\
\textbf{our + AttU-Net} & 93.63 & 91.09 & 91.59 & 98.20  \\
& \textcolor{red}{(↑4.05)} &  \textcolor{red}{(↑3.20)} & \textcolor{red}{(↑6.03)} & \textcolor{red}{(↑2.91)} \\
\hline
TransUNet \cite{chen2021transunet} & 89.71 & 88.86 & 84.53 & 95.73  \\
\textbf{our + TransUNet} & 93.86 & 91.83 & 91.66 & 98.11  \\
& \textcolor{red}{(↑4.15)} &  \textcolor{red}{(↑2.97)} & \textcolor{red}{(↑7.13)} & \textcolor{red}{(↑2.38)} \\
\hline
\end{tabular}
\end{table}

\subsection{Ablation Study On Spatial Prior Network}

To investigate the impact of different SP-Net backbone models on overall network performance, we replaced the Spatial Prior Network in SPG-CDENet with three representative segmentation models: UNet (a CNN-based architecture), AttU-Net (a CNN with attention mechanisms), and TransUNet (a hybrid CNN and Transformer architecture). This experiment was conducted using a pre-trained U-Net model as the initial backbone for training. Then, we directly swapped the pre-trained U-Net with AttU-Net and TransUNet without further training, and the results were still impressive. The results are shown in Table \ref{tab:7} and Table \ref{tab:8}.

On the Synapse dataset, our proposed model achieves substantial improvements when using ROI prior information generated by the U-Net model. Specifically, the average DSC improves by 9.12\%, while the average HD decreases by 26.95. This demonstrates that incorporating ROI prior information significantly enhances segmentation performance. Furthermore, when the pre-trained model was replaced with AttU-Net and TransUNet—without further training—the performance remained excellent, the average DSC improves by 8.14\% and 8.54\%, and the average HD decreases by 22.79 and 19.06, highlighting the robustness and plug-and-play capability of our approach.

On the ACDC dataset, consistent improvements are also observed across all three SP-Net variants. The average DSC increases by 6.46\%, 4.05\%, and 4.15\%, respectively. This further validates that our model is highly effective and stable across different datasets. Overall, the proposed method consistently outperforms baselines across diverse SP-Nets and datasets, confirming its effectiveness, adaptability, and reliability.
% In general, incorporating the Organ-aware Prior Network consistently improves performance for both U-Net and AttU-Net backbones, demonstrating the robustness of our approach. Second, organs that are more challenging to segment in the backbone models, such as the pancreas and gallbladder, benefit from larger improvements, indicating that incorporating anatomical priors effectively facilitates the segmentation of small and difficult-to-localize structures. Overall, these results validate that SPG-CDENet not only enhances baseline model performance but also effectively addresses the core challenges in multi-organ segmentation, particularly improving small organ delineation, boundary accuracy, and generalization across diverse anatomical structures.

% These results show that our method does not heavily rely on the ROI prior information, demonstrating its versatility across different pre-trained model.

\section{Conclusion}
%  Our proposed model consistently outperforms previous methods across both Synapse dataset and ACDC dataset, demonstrating its superior segmentation capability. Furthermore, the results validate the effectiveness of incorporating prior knowledge into medical image segmentation. By introducing the Organ-aware Prior Network and a dual-encoder design, our model leverages prior anatomical information more effectively, enhancing the segmentation accuracy of each structural component. This highlights the potential of prior-guided learning to achieve more robust and anatomically consistent medical image segmentation.
In this work, we propose a novel SPG-CDENet that integrates a Spatial Prior Network and a crossing dual-encoder network. Experiments on the Synapse and ACDC datasets demonstrate that our model achieves superior qualitative and quantitative performance. Furthermore, we analyze the impact of prior networks with different architectures and performance levels on the final segmentation results. Experimental results demonstrate that our prior network shows strong generalization and plug-and-play capability, enabling easy replacement of different prior models without performance degradation.To further validate our design, we conduct an ablation study on various interaction strategies between the global and local encoders. The results show that the proposed SCA module achieves the best performance, confirming its effectiveness in leveraging prior information for segmentation. In future work, we plan to extend our method to 3D medical image segmentation tasks to further enhance its generalization and applicability.

\end{document}